\documentclass[twocolumn,superscriptaddress,longbibliography,aps,preprintnumbers]{revtex4-2}

\usepackage{graphicx,color}
\usepackage{xcolor}
\usepackage{bm,bbm,amsmath,amssymb,braket}
\usepackage{epstopdf}
\usepackage{relsize}
\usepackage{epstopdf}
\usepackage{comment}
\usepackage{soul}
\usepackage{float}
\usepackage{lipsum}
\usepackage[urlcolor=blue,colorlinks=true,citecolor=blue,
	linkcolor=red,pdfstartview={FitH},bookmarks=false]{hyperref}
\usepackage{MnSymbol}
\usepackage{stmaryrd}
\usepackage{mathtools}

\graphicspath{{Figures/}{./Figures/}{.}{./paper_figures/}}
\setstcolor{red}

\definecolor{purple}{rgb}{0.5, 0., 0.8}

\definecolor{persianblue}{rgb}{0.11, 0.22, 0.73}

\definecolor{darkred}{RGB}{182, 24, 24}

\begin{document}
\title{Postselection-free ballistic-diffusive transition in monitored spin chains}

\author{K. G. S. H. Gunawardana}\email{Email:korala.gunawardana@tuni.fi}
\affiliation{Computational Physics Laboratory, Physics Unit, Faculty of Engineering and Natural Sciences, Tampere University, P.O. Box 692, FI-33014 Tampere, Finland}
\affiliation{Helsinki Institute of Physics P.O. Box 64, FI-00014, Finland}
\author{Ali G.  Moghaddam}\email{Email: ali.moghaddam@aalto.fi}
\affiliation{Department of Applied Physics, Aalto University, 02150 Espoo, Finland}
\affiliation{Computational Physics Laboratory, Physics Unit, Faculty of Engineering and Natural Sciences, Tampere University, P.O. Box 692, FI-33014 Tampere, Finland}
\affiliation{Helsinki Institute of Physics P.O. Box 64, FI-00014, Finland}

\author{Teemu Ojanen}\email{Email: teemu.ojanen@tuni.fi}
\affiliation{Computational Physics Laboratory, Physics Unit, Faculty of Engineering and Natural Sciences, Tampere University, P.O. Box 692, FI-33014 Tampere, Finland}
\affiliation{Helsinki Institute of Physics P.O. Box 64, FI-00014, Finland}

\date{\today}

\begin{abstract}
We study spin and entanglement dynamics in spin-1/2 XXZ chains under periodic monitoring and show that this system exhibits two measurement-induced phase transitions: a steady-state entanglement phase transition similar to those in monitored quantum circuits and a ballistic-to-diffusive transition in 
transient dynamics. Specifically, we discover that at low monitoring rate, an initial configuration containing a domain wall $|\uparrow\uparrow\uparrow\ldots \downarrow\downarrow\downarrow\ldots\rangle$ spreads ballistically while, at large monitoring rates, the domain melting is diffusive. Extensive numerical simulations, supported by theoretical arguments, indicate that the ballistic-diffusive transition is intimately interlinked with the entanglement phase transition. In contrast to the entanglement phase transitions, which require exponentially complex postselection, the ballistic-diffusive transition can be observed without postselection and constitutes an experimentally accessible manifestation of the many-body Zeno effect.        

\end{abstract}

\maketitle

\begin{figure}[t]
    \centering
    \includegraphics[width=.99\columnwidth]{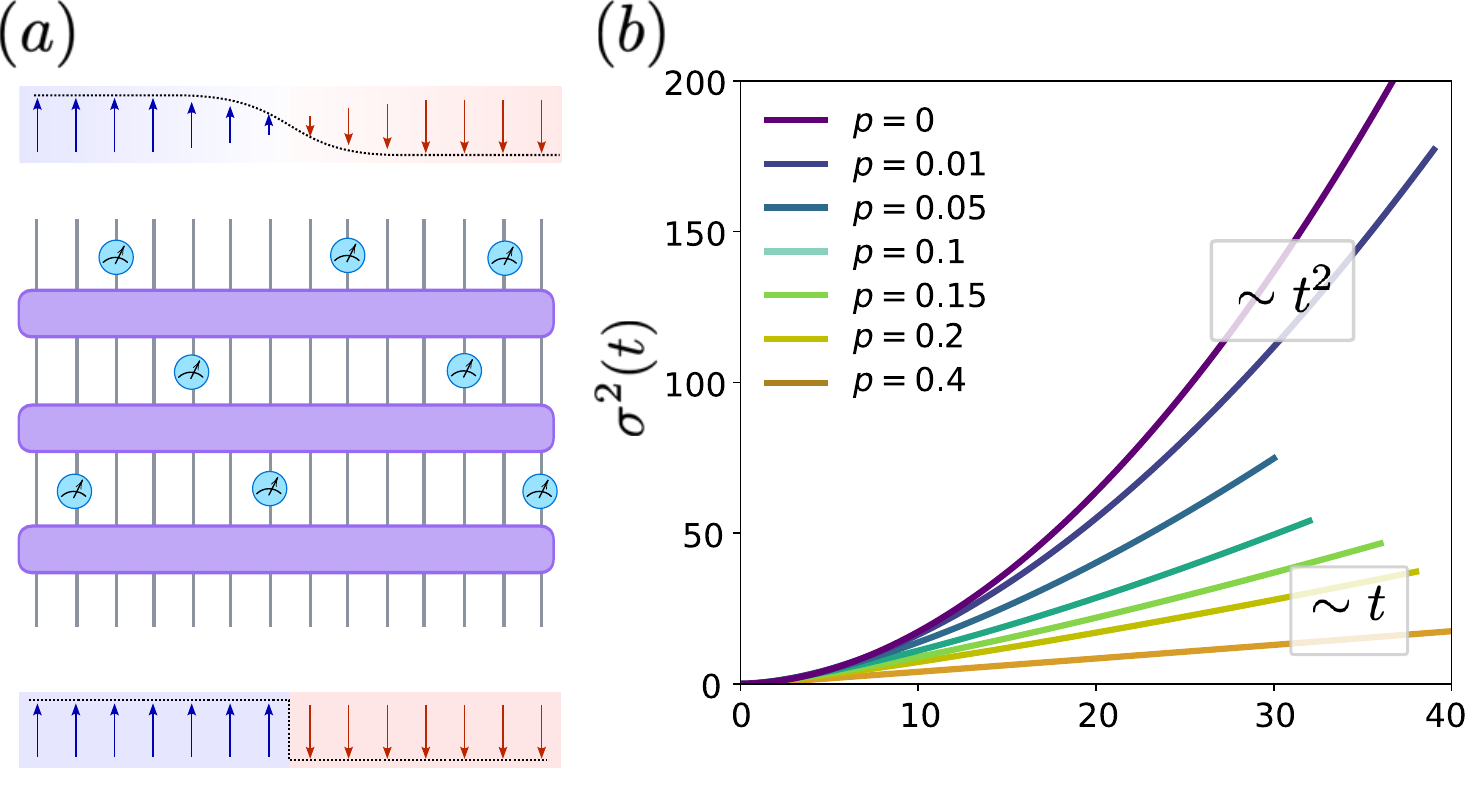}  
        \caption{ Ballistic to diffusive transition in periodically monitored spin chains. (a) Domain-wall initial state evolving under monitored dynamics. Unitary blocks (shown in purple), $U = e^{-iH\delta t}$, and measurement steps each correspond to a short time interval $\delta t \ll 1$. The probability of a measurement occurring at each measurement block is therefore $p_{\delta t} \sim p\,\delta t$. At later times, the domain wall melts, leading to a broader transition region between the fully up- and down-polarized domains.
        (b) Spreading of the initial sharp domain wall to a smoother profile quantified with a width $\sigma(t)$. At low monitoring rate $p<p_c$, the dynamics is ballistic ($\sigma\propto t$), while for $p>p_c$ it is diffusive ($\sigma\propto \sqrt{t}$).   }
    \label{fig1}
\end{figure}

\section{Introduction}

The interplay between unitary dynamics and measurement-induced non-unitarity represents a fundamental problem in quantum physics. Projective measurements collapse quantum states to eigenstates of the measured observable and profoundly alter the dynamics of isolated systems. This measurement back-action can suppress the unitary dynamics through the quantum Zeno effect, dramatically slowing or even freezing the natural evolution \cite{misra1977zeno,Wineland}. The study of \emph{monitored quantum systems} has experienced a renaissance in recent years, driven by advances on both theoretical and experimental fronts~\cite{
skinner_measurement-induced_2019,
li_quantum_2018,
li_measurement-driven_2019,
Fisher_random_quantum_circuits,vasseur2026houches,
potter_entanglement_2021,Monroe_measurement_2022}.
Theoretically, quantum information science has provided a framework to analyze how measurements affect many-body dynamics and entanglement \cite{Ludwig2020_measurement_criticality,bao_theory_2020,zabalo_critical_2020,lavasani_measurement-induced_2021,Altman2023,turkeshi_measurement-induced_2020,Mirlin2023,Buchhold_PRB}. Experimentally, quantum simulators and quantum computers now enable unprecedented control over many-body systems and direct observation of measurement outcomes of microscopic degrees of freedom \cite{google2023measurement,Minnich_2023measurement,XEB-experiment}. These capabilities have made quantum platforms ideal testbeds for exploring measurement-induced effects on quantum evolution.

Much of the recent activity has focused on measurement-induced phase transitions in monitored random circuits. While entanglement properties have been central to these studies, a diverse range of associated transitions has been identified, including purification, charge-sharpening, and transitions in magic (non-stabilizerness), among others \cite{gullans_dynamical_2020,Agrawal_2022,Halpern2023,Gullans2024magic,Zarand2025,choi_quantum_2020,Yamamoto2023}. These transitions reveal how the controllability of quantum states and access to microscopic information can fundamentally reshape many-body behavior, highlighting the central role of quantum information in modern condensed-matter physics \cite{zeng2019quantum,Augusiak2012,amico_entanglement_2008}. 

In this work, we study entanglement dynamics and transport in the spin-$1/2$ XXZ model 
\cite{Abanin2025,Subrahmanyam2004,Fazio2004,bayat2022entanglement,Gopalakrishnan_XXZ}
in the gapless regime under high-frequency monitoring, which corresponds to the discrete-time limit of continuous monitoring. This setup is illustrated in Fig. \ref{fig1}(a), where high-frequency monitoring is modeled by interrupting each short unitary evolution block $U=e^{-iH\delta t}$ with a measurement of each spin with probability $p_{\delta t} \sim p \delta t$ with $p$ corresponding to the measurement probability over a unit time scale ($\Delta t = 1 \gg \delta t$). Starting from a domain wall configuration, we analyze both the transient dynamics and the long-time properties after the system has relaxed to a steady state.

We find that, analogously to monitored random circuits, the system exhibits a phase transition between a volume-law and an area-law phase at a finite monitoring rate $\tilde{p}_c$. In addition to the steady-state entanglement phase transition, the melting of the domain wall reveals a ballistic-to-diffusive transition at a finite monitoring rate $p_c$. According to our numerical simulations, for $p < p_c$, the domain wall width expands ballistically, $\sigma(t) \propto t$, whereas for $p > p_c$ it grows only diffusively, $\sigma(t) \propto t^{1/2}$, as illustrated in Fig.~\ref{fig1}(b). Within the resolution of our finite-size numerics, both critical points $\tilde{p}_c$ and $p_c$ lie in the interval $0.1$--$0.2$. We argue theoretically that they should always satisfy $p_c \leq \tilde{p}_c$, and that the close vicinity of critical points is not accidental- both phenomena reflect how the qualitative behavior of local correlations is affected by the monitoring. Remarkably, in contrast to entanglement phase transitions, observing the ballistic-diffusive transition does not require preparation of a postselected measurement ensemble. Thus, the ballistic-diffusive transition also provides a way to probe the entanglement phase transition, which is exponentially difficult to observe directly.

\section{Model and Formulation}
\subsection{Periodically monitored spin chains}
We consider a one-dimensional spin-$1/2$ XXZ chain described by the Hamiltonian
\begin{equation}
H=\frac{J}{2} \sum_{i=1}^{L-1} \left(X_iX_{i+1}+Y_iY_{i+1}+\Delta Z_iZ_{i+1}\right),
\label{eq:Hamiltonian}
\end{equation}
where $J$ denotes the nearest-neighbor exchange coupling and $\Delta$ is the anisotropy parameter along the $z$ direction. Here, $X_i$, $Y_i$, and $Z_i$ are the Pauli operators acting on site $i$. We use dimensionless units where energy and time are measured in units of $J$ and $J^{-1}$, respectively. To further simplify the notation, we also set $J=1$ throughout the paper.
The Hamiltonian is invariant under rotations about the $z$ axis, which gives rise to a conserved $U(1)$ charge corresponding to the total magnetization, $[\sum_i Z_i,H]=0$. This conservation law is essential, as it makes spin transport a well-defined and meaningful probe of the many-body dynamics.

Our main focus is on the spin and entanglement dynamics of systems initially prepared in a product state. In particular, we consider the relaxation of domain-wall initial states of the form
\begin{equation}\label{eq:initial_state}
\ket{\psi(t=0)}=\ket{\uparrow \uparrow \cdots \uparrow \downarrow \downarrow \downarrow \cdots \downarrow },
\end{equation}
under monitored dynamics, where the natural unitary evolution is sequentially interrupted by measurements performed on a randomly selected subset of spins, as illustrated in Fig.~\ref{fig2}(a). We focus on the high-frequency monitoring regime, which asymptotically approaches continuous monitoring. In this setting, each short-time unitary evolution step $\hat{U}_{\delta t} = e^{i H\delta t}$ with $\delta t\ll 1$ is followed by a measurement layer applied with probability $p_{\delta t}$. As discussed in the next subsection, the unitary time-evolution operator $\hat{U}_{\delta t}$ is evaluated using the Trotter--Suzuki decomposition.

At each measurement layer, occurring immediately after a discrete time step $t=n\delta t$ (we denote the post-measurement time by $t^+=t+\epsilon$, with infinitesimal $\epsilon$, in order to distinguish it from the preceding unitary evolution), each spin is measured in the $z$ basis with probability $p_{\delta t}$ and left unperturbed with probability $1-p_{\delta t}$. For projective measurements, the corresponding measurement operators are $\hat{P}_{\uparrow}=\ket{\uparrow}\bra{\uparrow}$ and $\hat{P}_{\downarrow}=\ket{\downarrow}\bra{\downarrow}$, where $\ket{\uparrow}$ and $\ket{\downarrow}$ are eigenstates of the local $Z$ operator. In the simulations, measurement outcomes are sampled as random variables according to the Born probabilities for obtaining spin up or spin down, namely $|\bra{\uparrow}\psi(t)\rangle|^2$ and $|\bra{\downarrow}\psi(t)\rangle|^2$, respectively.

For a given stochastic realization of the system's evolution, at time $t=n \delta t$, a subset ${\cal M}_n = \{j_1,j_2,\cdots\}$ of sites is measured, yielding outcomes
${\bm m}_n \equiv (m_{j_1}, m_{j_2}, \cdots )$, where each $m_j\in\{\uparrow,\downarrow\}$. The corresponding post-measurement state can then be written as
\begin{equation}
\ket{\tilde{\psi}_{{\bm m}_n}(t+\epsilon)}=\bigotimes_{j\in {\cal M}_t} \hat{P}_{m_j}
\bigotimes_{j' \notin {\cal M}_t} \hat{\mathbbm 1}_{j'}
\ket{\psi (t)},
     \label{eq:Mstate}
\end{equation}
which must be normalized according to $\ket{\psi_{{\bm m}_n }(t+\epsilon)}=\ket{\tilde{\psi}_{{\bm m}_n }(t+\epsilon)}/|\ket{\tilde{\psi}_{{\bm m}_n }(t+\epsilon)}|$ in order to obtain a pure state dynamics of the state.

A complete sequence of measurement outcomes ${\bf m}=({\bm m}_1,{\bm m}_2, \cdots, {\bm m}_{N_T})$ over $N_T$ discrete time steps defines a particular quantum trajectory of the monitored system up to time $T=N_T\delta t$. In this discretized description of monitored evolution, for a fixed choice of measured sites throughout the evolution, denoted by ${\cal M}_{\rm tot}=\bigcup_{n=1}^{N_T} {\cal M}_n$, the total number of possible trajectories scales exponentially with the total number of measurement events. Formally, this gives $2^{|{\cal M}_{\rm tot}|}= 2^{\sum_{n=1}^{N_T} |{\cal M}_n| } $ possible outcome sequences. This exponential proliferation of trajectories immediately highlights a practical bottleneck, commonly referred to as the \emph{postselection problem} \cite{Ippoliti_postselection,McGinley2024,yamamoto2025postselection,Fazio2024postselection,Zhang2026postselection} in the context of monitored quantum dynamics, to which we will return in detail later.

\subsection{Numerical approach}

\begin{figure}[t]
    \centering
    \includegraphics[width=.8\columnwidth]{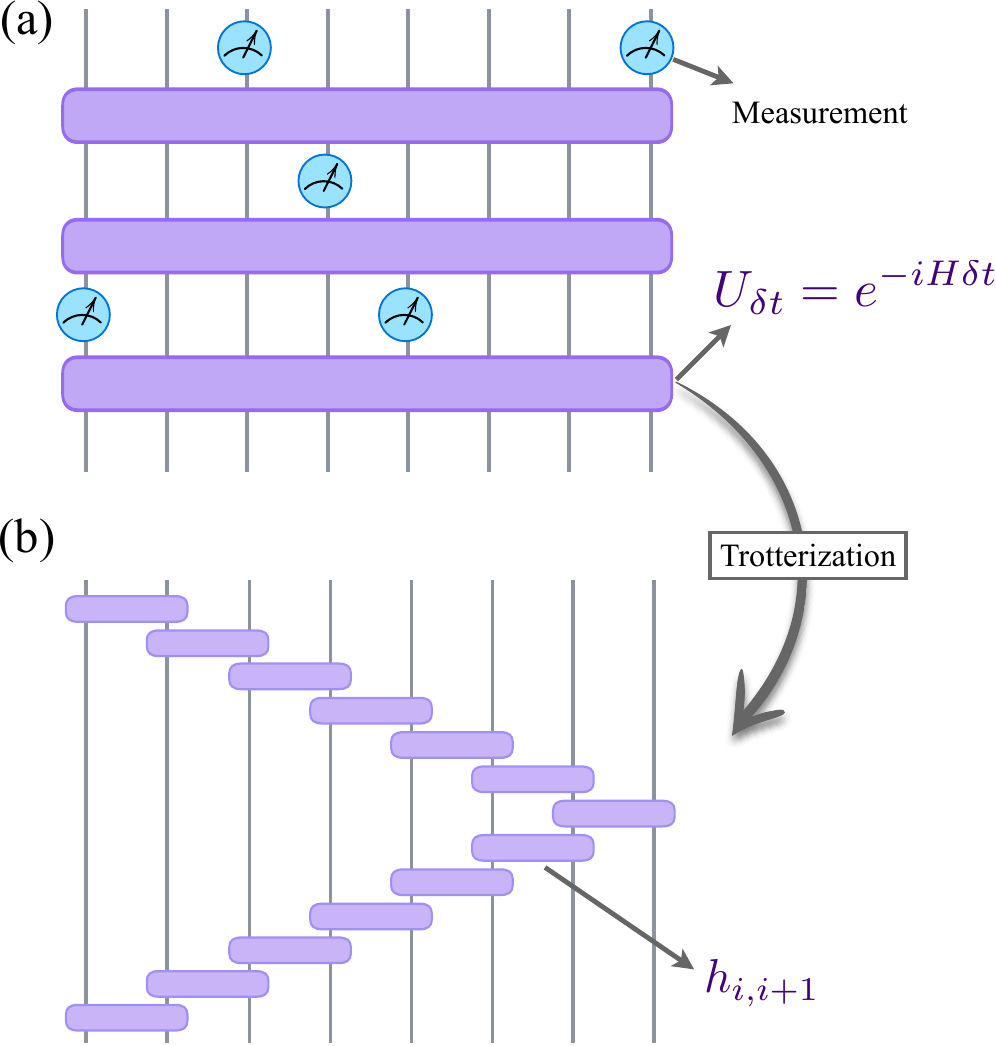}  
        \caption{Dynamics of spin chains in the high-frequency monitoring regime. (a): Temporal evolution consist of alternating unitary layers and measurement layers. (b): 2nd order  Trotter decomposition of the unitary layer. The two-qubit gates correspond to $e^{-i h_{i,i+1}\delta t/2 }$.   }
    \label{fig2}
\end{figure}

Starting from an initial product state $\ket{\psi(t=0)}$, we numerically simulate unitary evolution interrupted by measurements. We employ Matrix Product State (MPS) methods to efficiently represent the quantum many-body state and compute its time evolution. The MPS form of the quantum state can be written as \cite{Pollmann_MPS_2018,orus2014practical}
\begin{align}
     \ket{\psi}=\sum_{\{z_i\}}\sum_{\{\alpha_i\}} & u^{[1]z_1}_{l_1}u^{[2]z_2}_{l_1 l_2} \: \cdots \: u^{[L-1]z_{L-1}}_{l_{L-2}l_{L-1} }  u^{[L]z_L}_{l_{L-1} } \nonumber\\
     & \qquad \ket{z_1}\otimes\ket{z_2}\cdots \otimes \ket{z_L}, 
     \label{eq:mps}
\end{align}
where $\ket{z_i}$ are eigenstates of $Z_i$ and $u^{[i]z_i}_{l_{i-1} l_{i}}$ are rank-3 tensors at site $i$. In other words, $u^{[i]z_i}_{l_{i-1} l_{i}}$ are matrices with respect to the bond indices $\alpha_i$, but with a third index (denoted by the superscripts $z_i$) corresponding to the physical/site degree of freedom. The continuous unitary evolution is discretized into time steps $\delta t \ll 1$, as shown in Fig.~\ref{fig2}(a). The rectangles connecting the MPS legs represent the time-evolution operator over a single step, $\exp(-iH\delta t)\ket{\psi}$.

For the unitary update at each step $\delta t$, we use a second-order Trotter decomposition,
\begin{multline}
      e^{-iH\delta t}=e^{-ih_{12}\delta t/2}e^{-ih_{23}\delta t/2}\cdots e^{-ih_{L-1,L} \delta t/2}.e^{-ih_{L-1,L} \delta t/2} \\
      \cdots e^{-ih_{23}\delta t/2}e^{-ih_{12}\delta t/2} + O(\delta t^3).
      \label{eq:trotter}
\end{multline}
where the gates $h_{i,i+1}$ encode the local interaction between neighboring spins, such that the Hamiltonian \eqref{eq:Hamiltonian} can be expressed as
\begin{equation*}
    H=\sum_{i=1}^{L-1} h_{i,i+1}.
\end{equation*}
The effective quantum circuit corresponding to Eq.~\eqref{eq:trotter} is shown in Fig.~\ref{fig2}(b). We contract the decomposed unitary operator with the MPS to obtain the updated state at each time step. Numerical calculations are performed using the ITensorMPS package in the Julia programming language \cite{itensor,itensor-r0.3}.

As mentioned earlier, we consider periodic  monitoring in the high-frequency regime, i.e., a discrete-time version of continuous-time monitored dynamics. Assuming a measurement rate $\gamma$ in the continuous model, the measurement probability per discrete time step is $p_{\delta t}=\gamma \delta t$. This setup differs from monitored random quantum circuits, in which each (typically Haar-random) unitary layer is followed by a measurement layer: here, each unitary layer is close to the identity, and both the unitary and measurement steps occur on a time scale $\delta t \ll 1$, whereas in random circuits both are naturally defined on a unit time scale.

To connect with random monitored circuits, we match the probability that a given site remains unmeasured over a unit time interval $\Delta t=1$ in the two models. Equivalently, we identify one unit-time block of our discretized evolution with one cycle of unitary and measurement layers in the random-circuit setting. In the latter case, the probability of remaining unmeasured is $1-p$. In our discretized dynamics, a unit time interval contains $1/\delta t$ steps, and at each step the site remains unmeasured with probability $1-p_{\delta t}$; hence the probability of remaining unmeasured over $\Delta t=1$ is $(1-p_{\delta t})^{1/\delta t}$. Therefore, the effective measurement probability over a unit time $\Delta t=1$ in the frequent-monitoring model is
\begin{equation} \label{eq:p_versus_pdt}
    p \equiv 1-(1-p_{\delta t})^{1/\delta t}.
\end{equation}
For sufficiently small $\delta t$ and finite monitoring rate $\gamma \lesssim 1$ (so that $p_{\delta t}\ll 1$), this reduces to $p \approx p_{\delta t}/\delta t \equiv \gamma$. Equation~\eqref{eq:p_versus_pdt}, however, is more general and remains valid even at high monitoring rates, where $\gamma \gg 1$ and thus $p_{\delta t}\sim 1$. In the remainder of the paper, we report results in terms of $p$ using the substitution given by Eq. \eqref{eq:p_versus_pdt}.

\subsection{Postselection vs. no postselection}\label{subsec:observ}

A fundamental principle of quantum mechanics is that estimating the expectation value of an observable in a pure state requires measurements on many identically prepared copies of that state. In monitored dynamics, however, the central challenge is not simply the need for repeated state preparation, but the exponentially large number of possible quantum trajectories. With $\mathcal{M}$ mid-circuit measurements, each yielding one of multiple outcomes, there are exponentially many (e.g., $\sim 2^{\mathcal{M}}$) distinct measurement records $\mathbf{m}$, each corresponding to a different pure state $\ket{\psi_{\bf m}}$. Consequently, the probability of obtaining any specific trajectory is exponentially small in $\mathcal{M}$. To collect even a few samples of a given trajectory that is enough to estimate some property of $\ket{\psi_{\bf m}}$, one must repeat the entire experiment an exponential number of times. This exponential overhead is the essence of the \emph{postselection problem}.

Unfortunately, many key signatures of monitored dynamics and measurement-induced phase transitions appear only in quantities that are sensitive to the details of individual trajectories. Specifically, postselection is required for any quantity that is a nonlinear function of the single-trajectory density matrix. A central example is the von Neumann entanglement entropy,
\begin{align}
 S_\text{vN}(\ket{\psi_{\bf m}}) = - {\rm Tr}\big[ \rho_{{\bf m},A} \ln \rho_{{\bf m},A} \big],
\end{align}
where the reduced density matrix $\rho_{A}$ is obtained by tracing out the complement of subsystem $A$: $\rho_{{\bf m},A} = {\rm Tr}_{\bar{A}}(\ket{\psi_{\bf m}}\bra{\psi_{\bf m}})$. Because the entropy is a nonlinear function of 
the density matrix $\rho_{{\bf m},A}$, it cannot be reconstructed from trajectory-averaged linear observables; instead, one must first obtain results for individual trajectories, each of which is exponentially rare.

The trajectory-averaged entanglement entropy is computed as
\begin{align}\label{eq:traj_avg_S}
\overline{S_\text{vN}(t)}= \sum_{\bf m \in \{\pm 1\}^{\cal M} } p_{\bf m}  \: S_\text{vN}\big[\ket{\psi_{\bf m}(t)}\big],
\end{align}
where $p_{\bf m}$ is the probability of obtaining a given trajectory ${\bf m}$. Crucially, postselection remains necessary even for this averaged quantity: although one eventually averages over trajectories, each term in the sum requires access to a specific trajectory's pure state, and obtaining even a modest number of repetitions for each such trajectory demands an exponential number of experimental runs. In other words, for nonlinear functions such as the entropy, trajectory averaging does not commute with the function itself. The trajectory-averaged entropy $\overline{S_{\text{vN}}(\rho_A)}$ in Eq. \eqref{eq:traj_avg_S}
is not equal to the entropy of the trajectory-averaged density matrix $S_{\text{vN}}(\overline{\rho}_A)$, where 
\begin{align}\label{eq:mixed}
\overline{\rho}_A = \sum_{\bf m \in \{\pm 1\}^{\cal M} } p_{\bf m}  \: \rho_{{\bf m},A}.    
\end{align}

In numerical simulations, this difficulty is circumvented entirely: for each sampled trajectory, the full quantum state $\ket{\psi_{\bf m}}$ is directly accessible, allowing any observable or nonlinear functional (such as the reduced density matrix and its entropy) to be computed exactly without repetition. By contrast, in experiments, the quantum state can only be probed via projective measurements, which yield a single outcome per run; extracting even partial information about $\rho_{{\bf m},A}$ for a specific trajectory $\mathbf{m}$ thus requires repeating the experiment until that exponentially rare trajectory recurs sufficiently many times. In numerics, one needs only sample a modest number of trajectories to estimate trajectory-averaged quantities, whereas in experiments the exponential cost of postselection is unavoidable.

In contrast, for trajectory-averaged expectation values of observables,
\begin{align} \label{eq:Cpost}
 \overline{C(t)} = \sum_{\mathbf{m} \in \{\pm 1\}^{\cal M}} p_{\mathbf{m}} \bra{\psi_{\mathbf{m}}(t)} \hat{C} \ket{\psi_{\mathbf{m}}(t)},
\end{align}
postselection is not required at all, including in the experimental settings, despite what the expression might suggest at first glance. To see this, consider the expectation value of the same observable with respect to the trajectory-averaged (mixed) density matrix,
\begin{align}\label{eq:Cnopost}
\langle C(t)\rangle = \mathrm{Tr}_{A} \left[ \overline{\rho}_A  \: \hat{C}\right] = \mathrm{Tr}_A \Big[ \sum_{\bf m \in \{\pm 1\}^{\cal M} } p_{\bf m}  \: \rho_{{\bf m},A}(t) \: \hat{C}\Big] ,
\end{align}
Because the trace is linear, the expression can be re-written as 
\begin{align}
   \langle C(t)\rangle  = 
\sum_{\bf m \in \{\pm 1\}^{\cal M} } p_{\bf m}  \:  \mathrm{Tr}_A   \big[   \rho_{{\bf m},A}(t) \: \hat{C}\big],
\end{align}
which immediately implies that the two expressions \eqref{eq:Cpost} and \eqref{eq:Cnopost} coincide, since $\mathrm{Tr}_A   \big(   \rho_{{\bf m},A} \: \hat{C}\big) = 
\bra{\psi_{\mathbf{m}}(t)} \hat{C} \ket{\psi_{\mathbf{m}}(t)}
$ for a subsystem observable $\hat{C}$. The key insight is that any observable that is a linear function of the density matrix can be evaluated directly from the trajectory-averaged mixed state, without requiring access to individual trajectories. Experimentally, this means one can simply run the monitored dynamics, measure the observable $\hat{C}$ at the end of each run, and average the outcomes over many runs, without ever needing to reproduce any specific trajectory.

\subsection{Domain-wall spreading dynamics}

We investigate the dynamics of local magnetization starting from a domain-wall initial state. The spreading of the domain wall is extracted from the local magnetization density $\langle Z_i \rangle(t)$, computed using the density matrix in Eq.~\eqref{eq:mixed}. We consider initial product states with a sharp domain wall, as specified in Eq.~\eqref{eq:initial_state}, consisting of $L_{\uparrow}$ spins pointing up followed by $L_{\downarrow}$ spins pointing down. 
We take $L_{\downarrow}=L_{\uparrow}=L/2$ throughout.
To characterize the domain-wall profile, we define the normalized distribution function for the up spin density as
\begin{equation}
f(i,t)
=
\frac{\langle Z_i \rangle(t)+1}{2L_{\uparrow}} = 
\frac{\langle Z_i \rangle(t)+1}{L}.
\end{equation}
By virtue of the $U(1)$ symmetry of the magnetization, this distribution satisfies the conservation $\sum_{i} f(i,t)
=\sum_{i} f(i,t=0)=1$.
\par
In the absence of monitoring, the unitary evolution mixes the two magnetic domains, causing the sharp domain wall to spread and the magnetization profile to gradually homogenize. However, as we shall see, quantum measurements and monitoring substantially obstruct and slow down this melting process. Thus, the domain-wall spreading dynamics provides a natural probe of the interplay between unitary evolution and measurement-induced effects.
To quantify domain-wall spreading, we introduce the spreading function
\begin{align}
\sigma^{2}(t)
=
\tilde{\sigma}^{2}(t)-\tilde{\sigma}^{2}(t=0),
\label{eq:variance}
\end{align}
where variance $\tilde{\sigma}^{2}(t) =\langle i^2\rangle
-\langle i\rangle^ 2$ 
is calculated over the distribution $f(i,t)$.
As shown in Appendix~\ref{app:1}, the quantity $\sigma^{2}(t)$ effectively quantifies the domain-wall spreading and in the unmonitored limit and the regime of strong monitoring, it shows ballistic and diffusive signature, respectively.
\par
In general, we expect that asymptotic long-time behavior is of the form
\begin{align}\label{eq:diffusive}
\sigma^2(t) \propto t^\alpha.
\end{align}

\begin{figure*}[htp]
    \centering
\includegraphics[width=2\columnwidth]{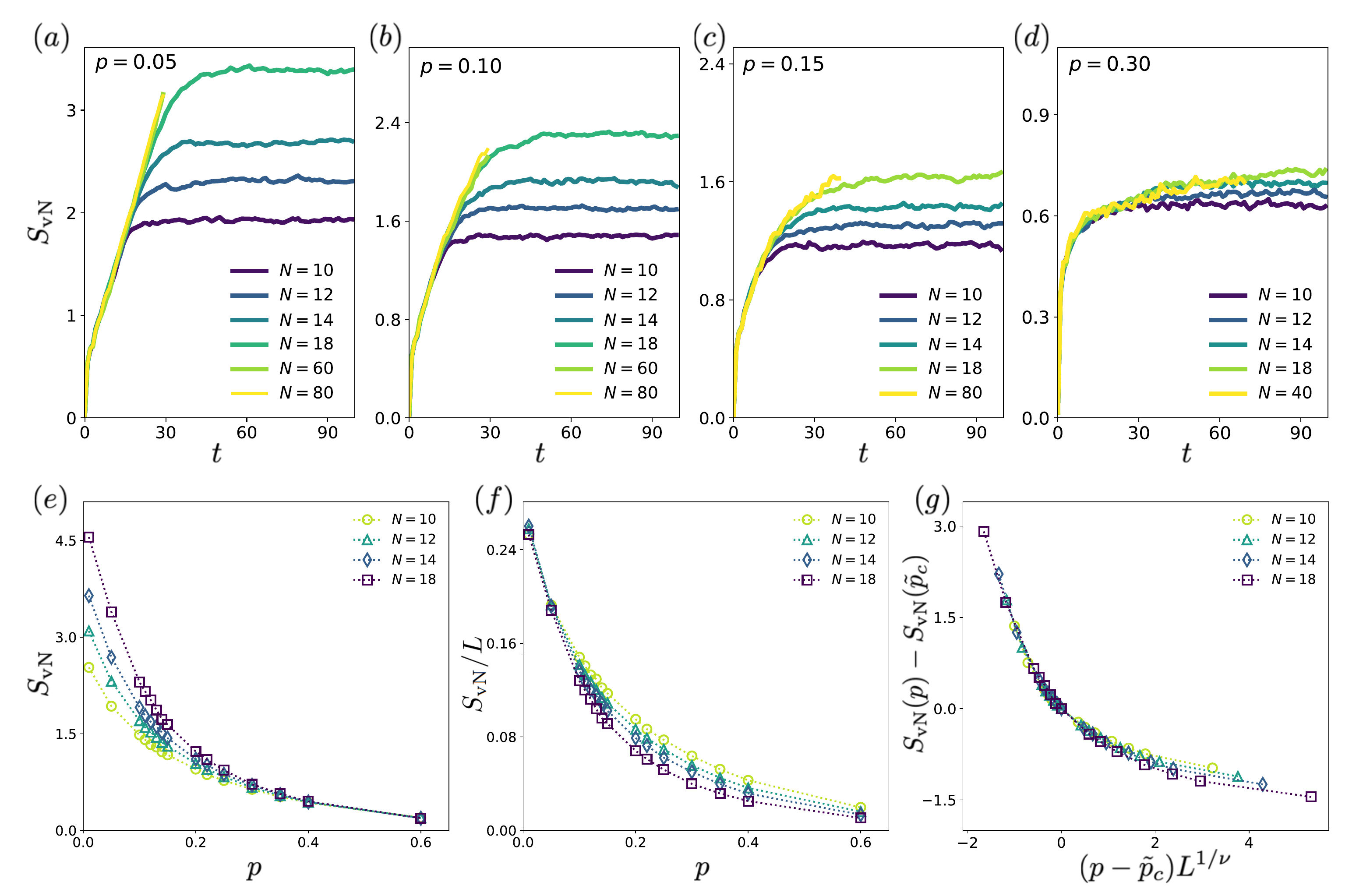} 
        \caption{Entanglement dynamics and entanglement phase transitions. (a)-(d): Entanglement entropy as a function of time for different monitoring strengths $p$. The time-evolution has been calculated by employing time step $\delta t=0.1$. In the small $p$ regime we employ the maximum MPS bond dimension $\chi =350$, while for $p>0.2$, the convergence of the entanglement entropy is reached when $\chi \approx 50$. (e)-(g): Trajectory-averaged entanglement entropy as a function of p. The results are obtained by averaging  over $500$ random trajectories at each $p\neq 0$. Additionally, the results have been averaged over the time window $60<t<100$ where the average value has been saturated. }
    \label{fig:entropy}
\end{figure*}

The cases $\alpha=2$ and $\alpha=1$ correspond to ballistic  and diffusive motion while non-integer values imply anomalous diffusion. As usual, extracting the asymptotic behavior from numerical data is obscured by a finite-time crossover dynamics. This is particularly challenging in monitored systems, which requires averaging over a large number of individual trajectories, making long-time studies computationally expensive. Thus, a direct fitting of the numerical results for $\sigma^2(t)$ using the Ansatz \eqref{eq:diffusive} is not a reliable strategy. Instead, to assess the asymptotic behavior accurately, it is convenient to study how the derivatives  $\frac{d(\sigma^2)}{dt}$ and $\frac{d^2(\sigma^2)}{dt^2}$ approach the asymptotic regime. The ballistic regime can be defined by $\frac{d^2(\sigma^2)}{dt^2}\to a>0$. The diffusive regime, in turn, corresponds to conditions $\frac{d^2(\sigma^2)}{dt^2}\to 0$ and $\frac{d(\sigma^2)}{dt}\to a>0$. Even when the derivatives have not relaxed exactly to constant values, it is possible to distinguish the qualitatively distinct behavior of different regimes as seen below.

\section{Results}

\subsection{Transient entanglement dynamics and entanglement phase transitions}

To set the stage, we first analyze transient entanglement dynamics starting from a domain-wall state $\ket{\psi(0)}=\ket{\uparrow \uparrow \uparrow \cdots \downarrow \downarrow \downarrow \cdots}$ and the entanglement phase transitions in a steady. In this work, we focus on the gapless regime $\Delta<1$, as it is known to support ballistic motion in the absence of monitoring, and study how the monitoring modifies the many-body dynamics.

In Fig.~\ref{fig:entropy} (a)-(d), we illustrate the build up of the entanglement entropy starting from a domain-wall state for different monitoring rates $p$. At small measurement rates $p<0.15$, the initial slope is roughly linear and independent of the system size. After a time $t\propto L$, a trajectory-averaged entropy saturates to a plateau, height of which is approximately proportional to the systems size. These features together indicate that the propagation of entanglement for small $p$ is ballistic and the steady-state entropy follows a volume-law scaling $\overline{S_\text{vN}(t)}\propto L$. As seen in Fig.~\ref{fig:entropy} (c), at $p=0.15$, the linear short-time behavior as well the size-dependent plateau scaling is no longer clearly resolved and for $p=0.3$, depicted in Fig.~\ref{fig:entropy} (d), we observe that the entropy saturation time and the asymptotic value are very weakly system-size independent. This is consistent with an entanglement area-law behavior in the steady state. 

The transient dynamics suggests that, in analogy to $U(1)$ symmetric monitored random circuits, the steady state undergoes a measurement-induced entanglement phase transition between a volume-law and an area-law phases for some measurement rate $0.1<p<0.2$. Indeed, the steady-state entropy as a function of the measurement rate, shown in Figs.~\ref{fig:entropy} (e)-(g), is reminiscent to that of monitored $U(1)$ symmetric circuits \cite{Agrawal_2022,Chakraborty2024}. The high $p$ regime $p\geq 0.3$, as seen in Fig.~\ref{fig:entropy} (e), follows very accurately an area-law scaling with size-independent entropy. In contrast, the low $p$ regime $p\geq 0.1$, follows accurately a volume-law scaling as indicated by  Fig.~\ref{fig:entropy} (f). 

The entanglement entropy data can be collapsed with a reasonable accuracy by a single-parameter scaling Ansatz $S_\text{vN}(p,L)-S_\text{vN}(\tilde{p}_c,L)=f[(p-\tilde{p}_c)L^{1/\nu}]$, using $\tilde{p}_c= 0.15 $ for the critical rate and $\nu= 1.2$ for the critical exponent. These values should be regarded as rough qualitative estimated, as the available system sizes limit more sophisticated analysis. This is a general difficulty in complex many-body systems, as analyzing systematic corrections to the single-parameter scaling become meaningful only for substantially larger system. It's notable, however, that the exponent $\nu$, which is expected to be determined only by symmetry and dimensionality, appears consistent with those observed in the monitored $U(1)$ random circuits \cite{Agrawal_2022,Chakraborty2024,moghaddam2023,Oshima2023U1}. Despite the fact that numerics can only offer a rough qualitative information on the critical parameters, the evidence supporting the existence of a measurement-induced entanglement phase transition is on par with generic unitary circuits.

\subsection{The ballistic-to-diffusive transition in magnetization dynamics}

We now turn to the domain-wall melting dynamics, characterized by the time evolution of the spatial variance $\sigma^2(t)$ defined in Eq.~\eqref{eq:variance}. We consider domain-wall initial states with equal-sized domains. The system size and propagation time are chosen such that opposite spin densities remains negligible at the chain boundaries, ensuring that results are insensitive to boundary effects and independent of the computational system size.

\begin{figure*}[htp]
    \centering
    \includegraphics[width=2.\columnwidth]{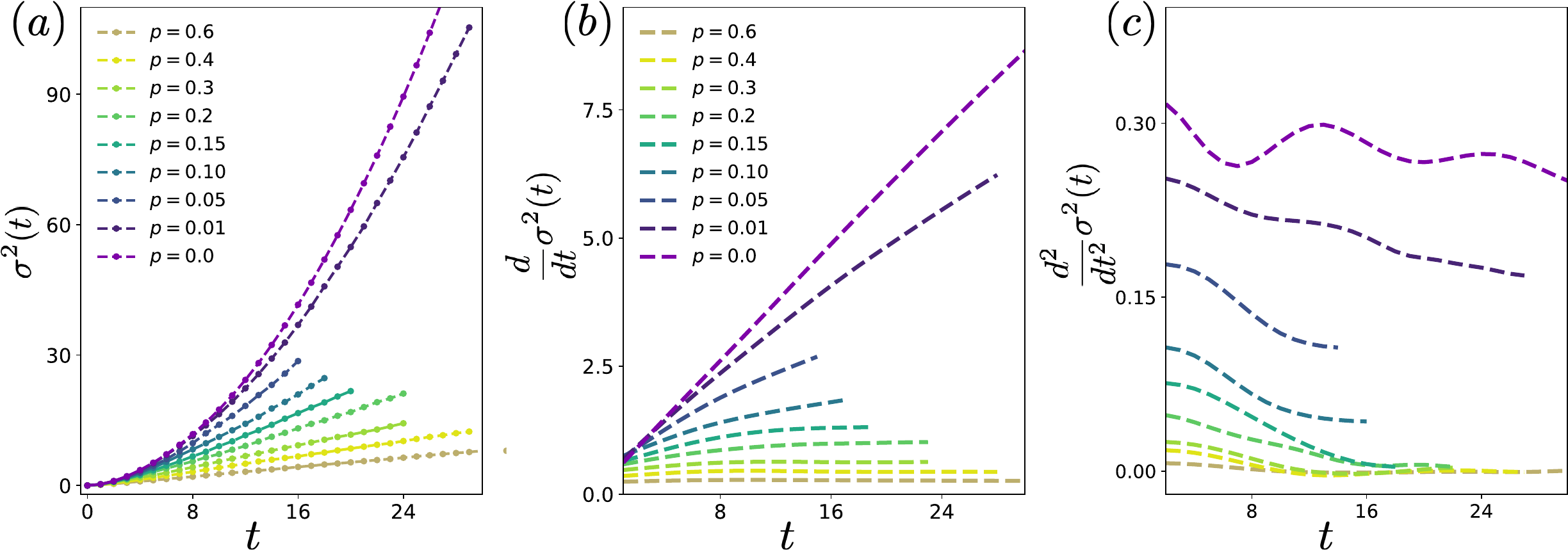}  
        \caption{Temporal spreading of the spin up domain for different monitoring strengths. (a): Dots correspond to the actual data points and the dashed line represents the spline fitting. (b): Derivative has been obtained by differentiating the spline fit in (a). (c): Second derivative is obtained by differentiating a spline-smoothed first derivate curve.    }
    \label{fig:spreading}
\end{figure*}

Results for $\sigma^2(t)$ at different measurement rates are presented in Fig.~\ref{fig:spreading}(a). Simulations in different parameter regimes present distinct computational tradeoffs. In the weak monitoring limit ($p\lesssim 0.1$), rapid entanglement growth increases the MPS computational cost; however, statistical fluctuations between quantum trajectories are suppressed, and convergence is achieved with relatively modest sampling ($\sim 40$ trajectories). In the strong monitoring regime ($p\gtrsim 0.2$), entanglement remains modest, reducing computational overhead; nevertheless, trajectory fluctuations are pronounced, requiring substantially larger statistical sampling ($\sim 1500$ trajectories at $p=0.6$). The intermediate regime ($0.1\lesssim p\lesssim 0.2$) exhibits moderate entanglement coupled with significant fluctuations, combining the challenges of both limits.

For the weak and intermediate regimes, we employ a system size of $L=160$ lattice sites. Simulations at $p=0$ and $p=0.01$ converge with 40 trajectory averages, while $p=0.6$ requires the maximum of 1500 trajectories. As the measurement rate increases, an increasing number of realizations becomes necessary to obtain smooth results. The trajectory averaging procedure constitutes the primary computational bottleneck in our calculations.

When the measurement rate is high enough $p\geq 0.15$, the long-time domain-wall spreading exhibits nearly linear behavior $\sigma^2(t)\propto t$, characteristic of diffusive dynamics. In contrast, for $p\leq 0.15$, the spreading is clearly superdiffusive. However, the precise nature of this transition is obscured by transient dynamics in the early-time regime, which is neither purely ballistic nor diffusive. Moreover, obtaining reliable long-time statistics through trajectory averaging becomes computationally prohibitive at intermediate measurement rates, as discussed above.

To extract the asymptotic behavior, we analyze the first and second time derivatives of $\sigma^2(t)$. Although the curves in Fig.~\ref{fig:spreading}(a) appear smooth, they inevitably contain noise from statistical fluctuations. Direct numerical differentiation is unstable in the presence of noise, so we instead fit a cubic spline to the data and compute derivatives from the fitted curve. This spline fitting acts as a regularization, smoothing high-frequency noise while preserving the qualitative long-time behavior. The results are presented in Fig.~\ref{fig:spreading}(b) and (c). The initial transient is clearly visible and explains why a naive power-law fit $\sigma^2(t)\propto t^\alpha$ over the accessible time window might not be sufficiently conclusive. For $p\geq 0.15$, the second derivative decays to zero while the first derivative remains positive, signaling diffusive dynamics $\sigma^2(t)\sim Dt$ with an effective diffusion constant $D$ that decreases with increasing $p$ but remains finite. In contrast, for $p\leq 0.1$, the second derivative relaxes to a positive constant, indicating ballistic behavior $\sigma^2(t)\propto t^2$.

The transition between these two regimes occurs near $p\sim 0.15$. For $p=0.1$, the second derivative has nearly reached its asymptotic value by the end of our simulation, whereas smaller $p$ values have not yet fully relaxed. However, on physical grounds we expect the curves for different $p$ to not intersect; thus, the second derivative for $p<0.1$ should eventually relax to values above that of $p=0.1$. This expectation is consistent with the known ballistic nature of unmonitored dynamics ($p=0$). We therefore conclude that all measurement rates in the range $0\leq p\leq 0.1$ exhibit ballistic behavior with a critical second derivative that decreases as $p$ increases. While the precise upper boundary for ballistic regime cannot be determined numerically, our results clearly place $p=0.15$ in the sub-ballistic regime.
The qualitative change from ballistic to diffusive dynamics represents a genuine transition between distinct dynamical regimes. Remarkably, the critical measurement rate for this ballistic-to-diffusive transition coincides, within our numerical resolution, with the critical rate for the entanglement phase transition discussed above, suggesting an intimate connection between these two phenomena.

\begin{figure}[t]
    \centering
    \includegraphics[width=1.0\columnwidth]{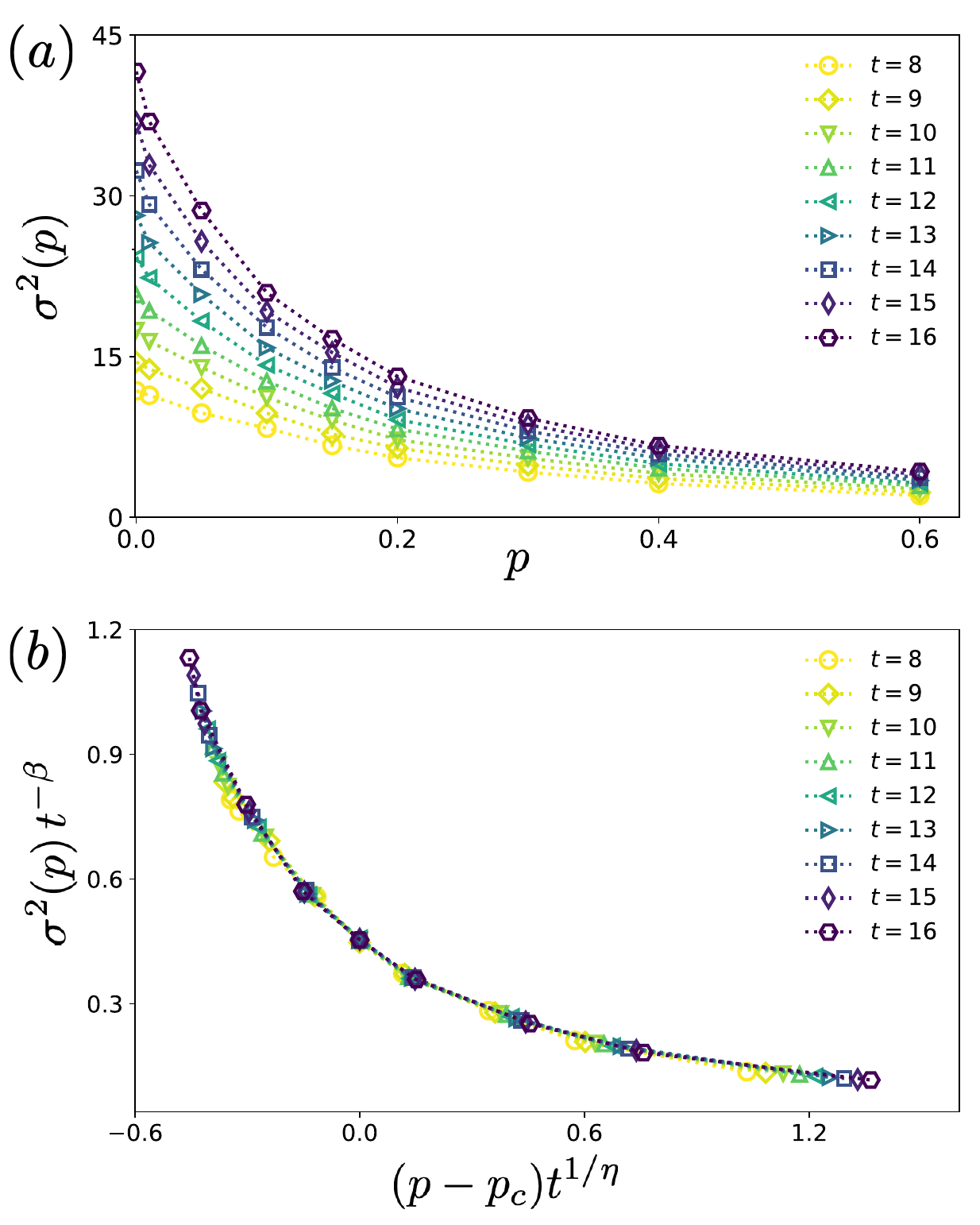}  
        \caption{Scaling analysis of the magnetization dynamics. (a): The effective spatial variance of the magnetization domain as a function of monitoring rate at different times. (b): Scaling collapse of the curves in (a) with the scaling form \eqref{eq:scalingform} and the critical parameters $p_c=0.15$, $\beta=1.3$ and $\eta=2.5$.    }
    \label{fig:scaling}
\end{figure}

An independent support for the measurement-induced ballistic-to-diffusive transition can be obtained through a scaling analysis. We postulate that the domain spreading obeys a scaling form 
\begin{equation}\label{eq:scalingform}
\sigma^2(p,t)=t^{\beta}F[(p-p_c)t^{1/\eta}],
\end{equation}
which describes a transition between two dynamical phases at a critical measurement rate $p_c$. Here, $\beta$ characterizes the domain-spreading exponent at criticality ($1\leq\beta\leq2$), and $\eta$ controls the correlation timescale $t_\circ\sim|p-p_c|^{-\eta}$ diverging at the criticality. As shown in Fig.~\ref{fig:scaling}, the numerical data approximately collapses onto a single curve, providing evidence for a ballistic-to-diffusive transition near $p_c\approx0.15$ with $\eta=2.5$ and $\beta=1.3$. These values are consistent with the critical point identified above.
We also note that the asymptotic form of function $F$ should satisfy 
\begin{align}
    F(x) \sim \left\{ 
    \begin{array}{cc}
         |x|^{\eta (2-\beta)} &  x\to - \infty\\
         x^{-\eta (\beta-1)} &  x\to + \infty
    \end{array}
    \right.
\end{align}

The extracted exponent $\beta=1.3$ characterizes anomalous diffusion at criticality and can be compared to predictions from the Kardar-Parisi-Zhang (KPZ) universality class \cite{KPZ-1986,barabasi1995fractal, Prosen2019, Yao_2022_KPZ}. For unmonitored spin chains at the Heisenberg point $\Delta=1$, KPZ scaling predicts a dynamic exponent $z=3/2$, giving $\sigma(p,t)\sim t^{2/3}$ and thus $\beta=4/3\approx1.33$. Our numerically extracted value $\beta=1.3$ agrees remarkably well with this KPZ prediction, suggesting that the  dynamics of domain wall melting at the criticality may belong to the KPZ universality class. However, given the limitations of the scaling collapse due to the finite time window, this conclusion should be viewed as suggestive rather than definitive.

In summary, we observe that domain-melting dynamics transitions from superdiffusive to diffusive behavior near $p_c\approx0.15$. Detailed analysis reveals a ballistic-to-diffusive phase transition whose critical point coincides, within our numerical resolution, with the steady-state entanglement phase transition identified earlier. Moreover, in App. \ref{app:1}, we model the domain-wall spreading using generalized hydrodynamics and derive the ballistic and diffusive limits analytically. In the unmonitored limit ($p=0$) the emergence of a light-cone structure \cite{Gopalakrishnan-2018,Gopalakrishnan-2025}, gives rise to $\sigma^2(t)\propto t^2 + {\cal O}\big[(1/L)^{2}\big]$. In the strong-monitoring limit, measurement-induced suppression of the dynamics dominates, leading to diffusive spreading $\sigma^2(t)\propto t + {\cal O}\big[(1/L)^{2}\big] $. These analytical predictions are consistent with the numerical observations presented above.

\section{Discussion}

Above, we established the existence of an entanglement phase transition along with a ballistic-diffusive transition in domain wall dynamics at a critical monitoring strength denoted by $\tilde{p}_c$ and $p_c$, respectively. Although accessible system sizes and time scales do not warrant a high-precision extraction of the critical points, both transitions occur within $0.1 < p_c, \tilde{p}_c < 0.2$. This agreement, within the resolution of the present study, is unlikely to be coincidental and instead suggests a deeper relationship between the two transitions. Ballistic domain melting at low measurement rates requires that certain operator correlations, which locally change the $U(1)$ charge, proliferate ballistically. This is necessary because coherent spin flips are required for a spin species to advance along the chain. In addition, the early-time linear growth of entanglement entropy and the linear scaling of saturation time with system size ($t \sim L$) in Figs.~\ref{fig:entropy}(a) and (b) indicate ballistically propagating quantum correlations in the volume-law phase. This linear scaling is broadly observed in random quantum circuits, suggesting a general connection between ballistic correlation propagation and the entanglement volume-law phase. 

In contrast, the entropy saturation time in the entanglement area-law phase is independent of system size, implying a finite correlation length. This finite length imposes an effective cutoff on coherent spin-flip processes, naturally yielding the diffusive, random-walk-like spreading of a domain wall in the entanglement area-law phase. Since local-charge-changing correlations are only a subset of all possible quantum correlations that contribute to entanglement entropy, one expects that ballistically spreading entanglement necessitates the ballistically spreading charge regime. Therefore, the critical points should satisfy $p_c \leq \tilde{p}_c$. Ultimately, these considerations suggest both transitions may manifest the same underlying phenomenon: a measurement-induced transition between a phase where quantum correlations proliferate ballistically and a phase with a constant correlation length. Whether the seemingly distinct aspects of the many-body quantum Zeno effect can be unified by this picture provides an interesting direction for future research.

As discussed in Sec.~\ref{subsec:observ}, a direct experimental observation of measurement-induced entanglement phase transitions requires postselection, which entails repeating the monitored evolution exponentially many times. In contrast, the $U(1)$ charge density is a property of a mixed ensemble of quantum trajectories, requires no postselection, and is thus straightforward to access experimentally. This raises the question how the average charge can be sensitive to the behavior of the entanglement entropy, which requires access to individual quantum trajectories. The resolution to this puzzle is that while observables linear in the density matrix are not sensitive to entanglement at a fixed time, tracking the transient evolution of the average charge reveals a collective change in the behavior of the trajectories $\{\ket{\psi_{{\bf m}}(t)}\}$. Because each individual trajectory $\ket{\psi_{{\bf m}}(t)}$ supports ballistically spreading correlations at low measurement rates, the charge dynamics calculated from the mixed density matrix \eqref{eq:mixed} also predict ballistic spreading. In the area-law phase, spatial correlations across all trajectories have a finite scale, meaning the charge dynamics of the ensemble must become sub-ballistic. Finally, unlike other charge-related transitions in monitored $U(1)$ symmetric circuits that rely on quantities non-linear in the density matrix \cite{Agrawal_2022,Chakraborty2024,poyhonen2024}, the ballistic-diffusive transition studied here uniquely manifests in the transient dynamics of a linear observable.

Finally, while we have focused on the continuous-time evolution of $U(1)$-symmetric spin chains, the phenomena analyzed here could be realized in $U(1)$-symmetric monitored quantum circuits as well. In fact, the time-discretized Trotter evolution illustrated in Fig. \ref{fig2}, could be regarded as one specific example where the unitaries on one layer are close to identity. Thus, the ballistic-diffusive transition can be studied on a existing quantum processors which implement mid-circuit measurement.

\section{Summary}

In this work, we studied transient dynamics in monitored $U(1)$ symmetric spin chains and discovered a phase transition which separates a ballistic charge motion at low measurement rates and a diffusive  dynamics at high measurement rates. The system also exhibits a measurement-induced entanglement phase transition, which is closely interlinked with the ballistic-diffusive transition. Observation of the ballistic-diffusive transition does not require a postselection and is straightforward to access compared to the exponentially complex entanglement phase transitions. Our work illustrates novel aspects of a many-body Zeno effect and introduces a practical experimental method to probe measurement-induced phase transitions.

\acknowledgements
This work was supported by the Finnish Research Council project 362573 and the Finnish quantum flagship program. This work is part of the Finnish Centre of Excellence in Quantum Materials (QMAT).

\appendix

\section{Analytical results for the domain wall spreading}\label{app:1}

It is straightforward to verify that the XXZ Hamiltonian is invariant under both the discrete spin-rotation operator $\mathcal{R}_x$ (which maps $Z_i \to -Z_i$) and the spatial parity operator $\mathcal{P}$ (which maps site $i \to -i$ relative to the center of the system). Consequently, the corresponding unitary evolution commutes with the composite symmetry operator $\mathcal{S} = \mathcal{P}\mathcal{R}_x$. While the initial domain wall state $|\psi_0\rangle$ explicitly breaks $\mathcal{R}_x$ and $\mathcal{P}$ individually, it respects the combined symmetry $\mathcal{S}$. Because this symmetry is preserved under the dynamics, the expectation value of the local magnetization must satisfy $\langle Z_i(t) \rangle = -\langle Z_{-i}(t) \rangle$, implying that the spatial profile of the $z$-magnetization remains strictly antisymmetric at all times. Furthermore, this symmetry is robust even in the presence of random monitoring. Provided that the measurement protocol is statistically symmetric, the ensemble-averaged local magnetization, $\mathbb{E}_{\{{\bf m}\}}[\langle Z_i(t) \rangle]$, remains invariant under $\mathcal{S}$.

It follows directly that the second moment, $\langle i^2 \rangle_t$, evaluated over the distribution $f(i,t) = \frac{1 + \langle Z_i(t) \rangle}{L}$, is entirely time-independent. Due to the antisymmetry of the magnetization profile, the dynamical term vanishes exactly, yielding:
\begin{align}
    \langle i^2 \rangle_{t} &= \sum_{i=-L/2}^{L/2} i^2 f(i,t) = \frac{1}{L} \sum_{i=-L/2}^{L/2} i^2 \big[1 + \langle Z_i(t) \rangle\big] \nonumber \\
    &= \frac{1}{L} \sum_{i=-L/2}^{L/2} i^2 = \frac{(L + 2)(L + 1)}{12} \approx \frac{L^2}{12}.
\end{align}
Notably, this exact cancellation holds equally well for the monitored evolution, provided one considers the trajectory-averaged magnetization. Therefore, the domain wall spreading function defined in the main text, actually reads
\begin{align}
    \sigma^2(t)=\tilde{\sigma}^2(t)-\tilde{\sigma}^2(0)
    = \langle i \rangle_{t=0}^2 - \langle i \rangle_t^2,
\end{align}
due the the cancellation of time-independent second moment term. Hence, the domain wall spreading function $\sigma(t)$
is basically governed by the first moment $\langle i \rangle_{t}$.

\subsection{Unmonitored limit: pure Hamiltonian dynamics}
In the unmonitored limit $p=0$, as established by generalized hydrodynamics, in the interacting, gapless regime ($|\Delta| < 1$), the XXZ model supports ballistic transport. Although the model is integrable and the exact magnetization profile can, in principle, be derived via the thermodynamic Bethe ansatz, its explicit analytical form is highly complex. Nonetheless, the macroscopic dynamics are strictly governed by a light cone $|i| \le vt$, where $v$ is the maximal dressed velocity of the magnons, which depends on the anisotropy $\Delta$. Inside this light cone, the profile assumes a smooth, dynamically scaled form; outside the light cone ($|i| > vt$), the domain wall strictly retains its initial step-function profile. 

Remarkably, we can extract the exact time-scaling behavior of $\langle i \rangle_t$ without requiring the explicit functional form of the magnetization inside the light cone. Approximating the discrete sum by an integral in the continuum limit ($i \to x$), the first moment evaluates to:
\begin{align}
    \langle i \rangle_t &\approx \langle x \rangle_t = \frac{1}{L} \int_{-L/2}^{L/2} x \big[1 + \langle Z \rangle(x,t)\big] dx \nonumber \\
    &= \frac{2}{L} \left[ \int_{0}^{vt} x \: \langle Z \rangle(x,t) dx - \int_{vt}^{L/2} x dx \right].
\end{align}
The integral over the region strictly outside the light cone evaluates trivially to $-\frac{L^2}{8} + \frac{v^2 t^2}{2}$. Inside the light cone, hydrodynamic scaling implies that the magnetization is a function only of the scaling variable $\zeta = x/t$, such that $\langle Z \rangle(x,t) = \mathcal{Z}(\zeta)$. Applying this substitution, the inside integral becomes:
\begin{equation}
    \int_{0}^{vt} x \: \mathcal{Z}(x/t) dx = t^2 \int_{0}^{v} \zeta \mathcal{Z}(\zeta) d\zeta \equiv -C_v t^2,
\end{equation}
where $C_v = -\int_{0}^{v} \zeta \mathcal{Z}(\zeta) d\zeta$ is a strictly time-independent, dimensionless constant. Recombining these terms, we obtain the final exact scaling behavior:
\begin{equation}
    \langle x \rangle_t = -\frac{L}{4} + \frac{2}{L} C'_v t^2,
\end{equation}
with $C'_v = \frac{v^2}{2} - C_v$ corresponding to the ballistic spreading in the unmonitored limit $p=0$.

Consequently, we get the following form for the dynamical spreading of the domain wall
\begin{align}
    \sigma(t) &= \sqrt{C'_v t^2 - \left(\frac{2 C'_v}{L}\right)^2 t^4} \nonumber \\
    &\approx \sqrt{C'_v} \: t \left[ 1 + \mathcal{O}\left(\frac{v^2 t^2}{L^2}\right) \right].
\end{align}
In the thermodynamic limit ($L \to \infty$), the extensive $\mathcal{O}(1/L^2)$ correction vanishes perfectly, revealing a strictly linear ballistic spreading $\sigma(t) \propto t$. For a finite chain, the approximate linear-in-time form remains highly accurate for early and intermediate times, provided the light cone is far from approaching the system boundaries ($vt \ll L$).

\subsection{Strong monitoring limit}

The framework established above can be extended at least in a qualitative manner to encompass the effects of local measurements. Under weak monitoring, the ballistic scaling structure is largely robust, and the domain wall continues to melt linearly in time. However, for sufficiently strong monitoring rates, the dynamics must undergo a significant change. The frequent projection of local spins and corresponding quantum Zeno effect effectively obstructs the ballistic transport of excitations. Consequently, the ballistic light cone $vt$ is suppressed and for strong enough monitoring we expect a diffusive envelope characterized by a dynamic length scale $\sqrt{D t}$, where $D$ represents an effective diffusion coefficient.

Hence, we can repeat the similar calculation in this regime by
replacing the ballistic boundary with the diffusive boundary $\sqrt{Dt}$ for the spreading of domain wall:
\begin{align}
    \langle x \rangle_t = \frac{2}{L} \left[ \int_{0}^{\sqrt{Dt}} x \:  \mathcal{Z}_d(x/\sqrt{t}) dx - \int_{\sqrt{Dt}}^{L/2} x dx \right].
\end{align}
Then, assuming that magnetization is now a function of the diffusive scaling variable $\xi= x/\sqrt{Dt}$, 
we find that the first moment is linear in time, as
\begin{equation}
    \langle x \rangle_t = -\frac{L}{4} + \frac{2}{L} C'_D t,
\end{equation}
where $C'_D = \frac{D}{2} - C_D$ and $C_D = -\int_{0}^{\sqrt{D}} \xi \mathcal{Z}_d(\xi) d\xi$. 
Therefore, the domain-wall spreading function for the strongly monitored regime reads
\begin{align}
    \sigma(t) &= \sqrt{C'_D t - \left(\frac{2 C'_D}{L}\right)^2 t^2} \nonumber \\
    &\approx \sqrt{C'_D\, t} \:   \left[ 1 + \mathcal{O}\left(\frac{D t}{L^2}\right) \right].
\end{align}

\bibliography{refs}
\end{document}